\documentclass{elsart}
\usepackage{amsmath}
\usepackage[english]{babel}
\usepackage{graphicx}

\begin{document}
\begin{frontmatter}

\title{Constraints on the Cosmological Parameters in the
Relativistic Theory of Gravitation}

\author{V. L. Kalashnikov\thanksref{l1}}
\thanks[l1]{This work was inspired by Prof. A.A.
Logunov noted the significance of the causality principle for the
cosmological models in RTG. Author appreciate him for the helpful
discussion. The work was carried out in the computer algebra
system Maple 6 \cite{11}. Author is Lise Meitner Fellow at
Technical University of Vienna (Project M611)}
\address{Institut f\"ur Photonik, TU Wien, Gusshausstr. 27/387,
A-1040 Vienna, Austria}
\ead{vladimir.kalashnikov@tuwien.ac.at}

\begin{abstract} The causality principle imposes the constraints on
the cosmological parameters in the relativistic theory of
gravitation. As a result, X-matter causes the quite definite
cosmological scenario with the alternate acceleration and
deceleration and the final recollapse
\end{abstract}

\end{frontmatter}

The relativistic theory of gravitation (RTG) denies the total
geometrization and is based on the traditional field approach
\cite{1,2}. The gravitation is interpreted as the tensor field
generated by the conserved energy-momentum tensor of the matter.
There is the background spacetime of the Minkowski's type, which
can be restored in any situation. This preserves the unambiguous
physical content of the gravitational phenomena and simplifies the
unification of the views on the gravitation, on the one hand, and
the quantum mechanics, on the other hand.

It is known \cite{2}, that the usual cosmological solutions in RTG
do not agree with the modern observational data \cite{3,4} because
they predict the decelerated character of the cosmological
expansion at the present era. The insertion of the cosmological
term into field equation destroys the logical structure of the
theory because this requires to insert the additional repulsing
physical field, which is not affected by the matter.

The generalization of the field equation by the means of the
insertion of the scalar component with uniquely defined potential
allows the inflationary expanded solutions in RTG \cite{5}.
However, there is the purely phenomenological approach resulting
in the variety of the cosmological scenarios in the framework of
RTG, which is based on the modification of the matter
energy-momentum tensor \cite{6}. Such modification is produced by
the so-called "dark energy" (X-matter) term with the exotic
equation of state $p_x$=$w_x$$\rho_x$, where $w_x$$<$0 ($p_x$ and
$\rho_x$ are the pressure and the density, respectively).

However, Prof. A.A. Logunov kindly suggested \cite{7}, that
tacking into account the causality principle imposes the
constraints on the physically admissible solutions in the latter
approach. This allows to choose among the physically meaningful
cosmological parameters.

Here we will consider the aspects of the X-matter induced
cosmological evolution in RTG, which are inspired by the causality
principle. As a result, the defined class of the cosmological
scenarios will be selected, and the limits of the maximal scaling
factor as well as the approximated value of $w_x$ will be defined.

Let us begin with the usual assumption of homogeneity and isotropy
of the effective Riemannian spacetime produced by the action of
the gravitational field. The corresponding interval in the
spherical coordinates is \cite{2}:

\begin{equation}\label{1}
ds^2  = d\tau ^2  - \alpha a(\tau )^2 \left[ {dr^2  + r^2 \left(
{d\theta ^2  + \sin (\theta )^2 d\phi ^2 } \right)} \right],
\end{equation}

\noindent Here $\tau$ is the proper time, $a(\tau)$ is the scaling
factor; $\alpha$ is the constant of integration. This form of the
homogeneous and isotropic interval describing the globally flat
spacetime follows from the field equations, which have the form:

\begin{equation}\label{2}
G_n^m  - \frac{{m^2 }}{2}(\delta _n^m  + g^{mk} \gamma _{kn}  -
\frac{1}{2}\delta _n^m g^{pk} \gamma _{pk} ) =  - 8\pi T_n^m,
\end{equation}

\begin{equation}\label{3}
D_m \tilde g^{mn}  = 0,
\end{equation}

\noindent where $G_{n}^{m}$ is the Einstein's tensor defined on
the effictive Riemannian spacetime with the metrics $g^{mn}$;
$\gamma^{mn}$ is the metrics of the flat background Minkowski
spacetime, $D_{m}$ is the covariant derivative on the background
spacetime, $\tilde g^{mn} = \sqrt { - g} g^{mn}$, $c = G = \hbar =
1$, $m$ is the graviton's mass (the inverse transition to the
ordinary units corresponds to $m$$\rightarrow$$m c^2/\hbar$).

We choose the Galilean metrics as a background. The crucial
departure from \cite{6} is tacking into account the causality
principle in the framework of RTG \cite{8}: "the causality cone of
the effective Riemannian spacetime should be positioned inside the
causality cone of the Minkowski spacetime". As a result, the
arbitrary isotropic vector $u^m$ obeys:

\begin{equation}\label{4}
\gamma_{mn} u^m u^n = 0,
\end{equation}

\begin{equation}\label{5}
g_{mn} u^m u^n \leq 0,
\end{equation}

From the Eqs. (\ref{1}, \ref{4}, \ref{5}) we have the key
condition \cite{2}:

\begin{equation}\label{6}
a(\tau)^4 - \alpha < 0,
\end{equation}

\noindent which eliminates the cosmological solutions with the
eternal expansion and keeps the scenario of IV type in \cite{6}.
It is convenient to assign $\alpha$=$a_{max}^4$, where $a_{max}$
is the maximal value of the scaling factor. Then the cosmological
equations are:

\begin{equation}\label{7}
\left( {\frac{{\dot a}}{a}} \right)^2  = \frac{8}{3}\pi \rho (\tau
) - \frac{{m^2 }} {12}\left( {2 + \frac{1} {{a(\tau )^6 }} -
\frac{3} {{a(\tau )^2 a_{\max }^4 }}} \right),
\end{equation}
\begin{equation}\label{8}
\frac{{\ddot a}}{a} =  - \frac{4}{3}\pi \left( {3p(\tau ) +
\rho (\tau )} \right) - \frac{1}{6}m^2 \left( {1 - \frac{1}{{a^6 }}} \right).
\end{equation}

From the Eq. (\ref{7}) one can obtain the expression for the
minimal density of the matter corresponding to the maximal scaling
factor \cite{2}:

\begin{equation} \label{9}
\rho _{\min }  = \frac{{m^2 }} {{16\pi }}\left( {1 - \frac{1}
{{a_{\max }^6 }}} \right).
\end{equation}

Let's suppose $a_{max}$$\gg$$a_0$, where $a_0$ is the present
scaling factor. This assumption is suggested by the accelerated
expansion of the universe at the present era. Then the minimal
density is defined by the form of the matter with the slowest
decrease produced by the growing scaling factor. As
$\rho$$\propto$$a(\tau)^{-3(1+w)}$, the X-matter with $w_x$$<$0
dominates in the late universe. Hence the Eq. (\ref{9}) results
in:

\begin{equation}\label{10}
\frac{{\Omega _g }} {{\Omega _x }} \cong a_{\max }^{ - 3\delta },
\end{equation}

\noindent where $\Omega _g  = m^2 / (6 H_0^2)$ and $\Omega _x =
 8 \pi \rho _x /(3 H_0^2)$ are the density parameters for
the gravitons and the X-matter, respectively, $H_0$ is the Hubble
constant; $\delta$$=$$1+w_x$ is the deviation of the X-matter
state parameter from that for the pure cosmological constant.

The Eq. (\ref{7}) defines the modified cosmic sum rule:

\begin{equation}\label{11}
\Omega _r  + \Omega _m  + \Omega _x  - \frac{3} {2}\Omega _g
\left( {1 - \frac{1} {{a_{\max }^4 }}} \right) = 1
\end{equation}

\noindent where $\Omega _m  = 8 \pi \rho _m/(3 H_0^2)$ and $\Omega
_r  = 8\pi \rho _r/(3 H_0^2)$ are the density parameters for the
nonrelativistic and relativistic matter with $w$=0 and 1/3,
respectively. One can see, that the sum of the matter densities
exceeds the critical density due to the graviton's mass
contribution. Although the modern data demonstrates some exceeding
$\Omega_{tot}$=$\Omega_x$+$\Omega_m$+$\Omega_r$$\cong 1.11 \pm
0.07_{ - 0.12}^{ + 0.13}$ \cite{9}, we suppose that the gravitons
induced effect is too small to be revealed in these observations.

The parameters of the Eq. (\ref{10}) can be concretized
additionally by taking into account the accelerated expansion of
the universe at the present era and the observational data from
BOOMERANG, MAXIMA and COBE \cite{3,4,9}. The acceleration
parameter $q$ = $\left.{(d^2 a/d\tau ^2 )} \right|_0$/$(a_0 H_0^2
)$$\cong$0.33$\pm$0.17,$\Omega_m$$\cong$0.37 $\pm$ 0.07,
$\Omega_x$$\cong$0.71 $\pm$ 0.05.

From the Eqs. (\ref{7},\ref{8}) we have

\begin{equation}\label{12}
q = \frac{{\Omega _x \left( {1 - \frac{3} {2}\delta } \right) -
\frac{1} {2}\Omega _m  - \Omega _r }} {{\Omega _{tot}  - \frac{3}
{2}\Omega _g }}.
\end{equation}

\noindent If the gravitons and the relativistic matter do not
contribute in the present state, the combination of observational
data and Eq. (\ref{12}) results in the estimation of $\delta$:

\begin{equation}\label{13}
\delta  = \frac{2} {3}\left( {1 - q} \right) - \frac{{\Omega _m }}
{{3\Omega _x }}\left( {1 + 2q} \right) \cong 0.16_{ - 0.09}^{ +
0.11}.
\end{equation}

Now we have to estimate the maximal value of $\Omega_g$ tacking
into account the condition $a_{min}$$<$$a_r$, where $a_{min}$ is
the minimal scaling factor, $a_r$ is the scaling factor at the end
of the radiation dominating era \cite{6}. $a_{\min }$ $\approx$
$\sqrt {{\raise0.7ex\hbox{${\Omega _g }$} \!\mathord{\left/
 {\vphantom {{\Omega _g } {2\Omega _r }}}\right.\kern-\nulldelimiterspace}
\!\lower0.7ex\hbox{${2\Omega _r }$}}}$ results in
$\Omega_g$$<$10$^{-11.7}$. This condition in the combination with
Eqs. (\ref{10}, \ref{13}) gives Fig. 1. One can see, that
$log_{10}(a_{max})$=10$\div$55 (with the most probable value in
the vicinity of 14) and, it is natural, the approach of $w$ to -1
or $\Omega_x$ to 1 increases the maximal scaling factor due to
growing negative pressure of the X-matter.

\begin{figure}

\caption{\label{fig1} The logarithm of the maximal $\Omega_g$
versus the logarithm of the maximal scaling factor for
$\delta$=0.27 (1), 0.16 (2), 0.07 (3); ($\Omega_x$,
$\Omega_m$)=(0.66,0.44), (0.71,0.37), and (0.76,0.3),
respectively. The dashed curve is the maximal $\Omega_g$ resulted
from $a_{min}$$<$$a_r$.}

\centering\includegraphics{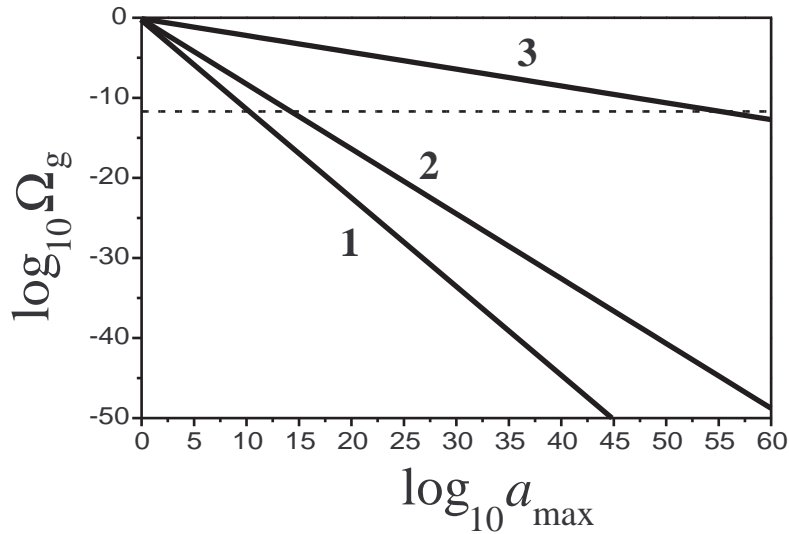}
\end{figure}

\begin{figure}

\caption{\label{fig2} The regions of the first deceleration for
$\delta$=0.27 (1), 0.16 (2), 0.07 (3); ($\Omega_x$,
$\Omega_m$)=(0.66,0.44), (0.71,0.37), and (0.76,0.3),
respectively. The black region is common for all parameters, the
end of the deceleration eras is different for the different
parameters (filled regions 1, 2 and 3).}

\centering\includegraphics{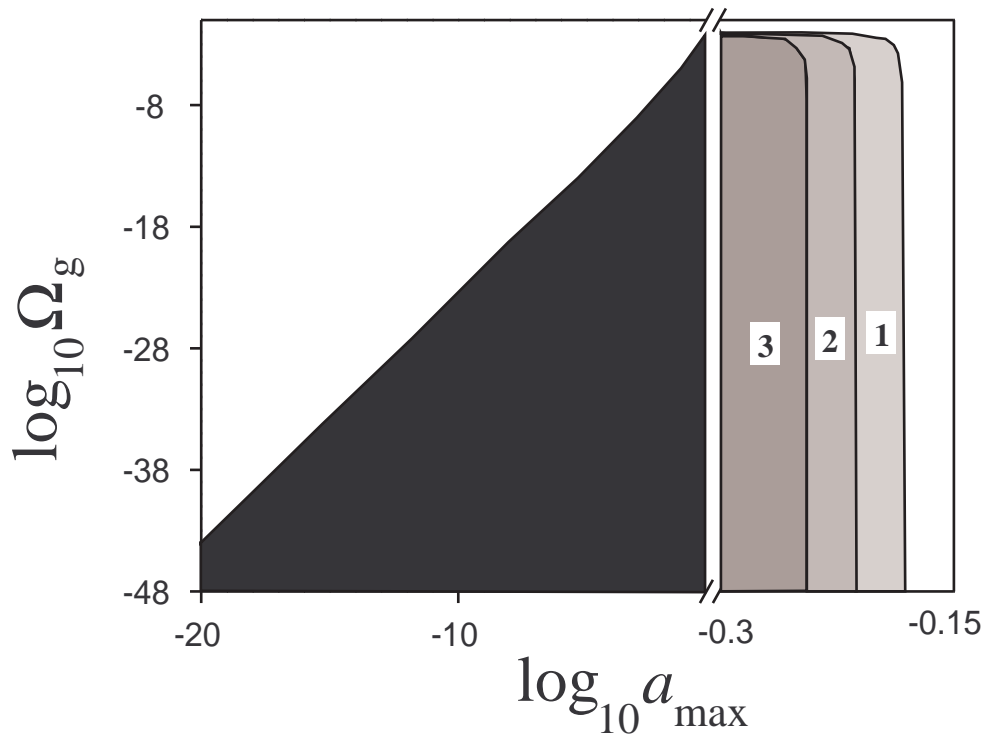}
\end{figure}

The regions of the accelerated (decelerated) expansion can be
found from Eq. (\ref{8}). The boundaries of these regions are
defined by the solutions of the following equation:

\begin{eqnarray}\label{14}
\Omega _m a^2  + 2\Omega _r a^2  - \left( {2 - \delta }
\right)\Omega _x a^{3(2 - \delta )}  +\\ \nonumber
 2\Omega _g \left( {a^6  -
1} \right) = 0.
\end{eqnarray}

The corresponding scenario has a complicated loitering character
(see \cite{6}): {\it acceleration} $\rightarrow$ {\it
deceleration} $\rightarrow$ {\it acceleration} $\rightarrow$ {\it
deceleration} $\rightarrow$ {\it recollapse}. We live at the era
of the second acceleration, which began not long before the
present time (see Fig. 2). A long first deceleration should have
the pronounced observational consequences, for instance, in the
large-scale structure formation. The second deceleration era and
the recollapse turning point are too remote from us, but the
estimated upper limit of the universe age is not too large in the
comparison with the so-called "dark era" representing the decay of
all known physical processes \cite{10}.

In the conclusion, the causality principle in the relativistic
theory of gravitation imposes the constraints on the cosmological
parameters defining the acceleration behavior of the universe at
the present time. The existence of the minimal and maximal scaling
factors requires to choose the fixed scenario with complicated
loitering behavior. For the observational values of the usual and
X- matter densities the deviation of the X-matter state from pure
vacuum one is $0.16_{ - 0.09}^{ + 0.11}$, which results in the
maximal scaling factor $\sim$ $10^{10}$$\div$$10^{55}$.

\end{document}